# GALACTIC INTERNET MADE POSSIBLE BY STAR GRAVITATIONAL LENSING

**Claudio Maccone**
Technical Director, International Academy of Astronautics, Via Martorelli 43, 10155 Torino (TO), Italy
clmaccon@libero.it and claudio.maccone@iaamail.org

**ABSTRACT**
A **Galactic Internet** may already exist, if all stars are exploited as gravitational lenses.
   In fact, the gravitational lens of the Sun is a well-known astrophysical phenomenon predicted by Einstein's general theory of relativity. It implies that, if we can send a probe along any radial direction away from the Sun up to the minimal distance of 550 AU and beyond, the Sun's mass will act as a huge magnifying lens, letting us "see" detailed radio maps of whatever may lie on the other side of the Sun even at very large distances. The 2009 book by this author, ref. [1], studies such future FOCAL space missions to 550 AU and beyond.
   In this paper, however, we want to study another possibility yet: how to *create* the future interstellar radio links between the solar system and any future interstellar probe by utilizing the gravitational lens of the Sun as a huge antenna. In particular, we study the Bit Error Rate (BER) across interstellar distances with and without using the gravitational lens effect of the Sun (ref. [2]). The conclusion is that only when we will exploit the Sun as a gravitational lens we will be able to communicate with our own probes (or with nearby Aliens) across the distances of even the nearest stars to us in the Galaxy, and that at a reasonable Bit Error Rate.
   We also study the *radio bridge* between the Sun and any other Star that is made up by the two gravitational lenses of both the Sun and that Star. The alignment for this radio bridge to work is very strict, but the power-saving is enormous, due to the huge contributions of the two stars' lenses to the overall antenna gain of the system. For instance, we study in detail:
1) The Sun - Alpha Cen A radio bridge.
2) The Sun – Barnard's Star radio bridge.
3) The Sun – Sirius A radio bridge.
4) The radio bridge between the Sun and any Sun-like star located in the Galactic Bulge.
5) The radio bridge between the Sun and a Sun-like star located inside the Andromeda galaxy (M31).
   Finally, we find the information channel capacity for each of the above radio bridges, putting thus a physical constraint to the amount of information transfer that will be possible even by exploiting the stars as gravitational lenses. The conclusion is that a GALACTIC INTERNET is indeed physically possible. May be the Galactic Internet is already in existence, and was created long ago by civilizations more advanced than ours. But this fact is being realized only in the year 2011 by Humans!

## 1. INTRODUCTION: THE SUN AND ALL STARS AS GRAVITATIONAL LENSES

   In a recent book (ref. [1]) this author studied the Sun as a gravitational lens and the relevant set of FOCAL (an acronym for "Fast Outgoing Cyclopean Astronomical Lens") space missions. According to general relativity, all stars are endowed with a powerful focusing effect called "gravitational lensing". This means that plane electromagnetic waves reaching the proximity of a (spherical) star from a distant radio source are deflected by the star gravity field and made to focus on the opposite side, as shown in Figure 1.

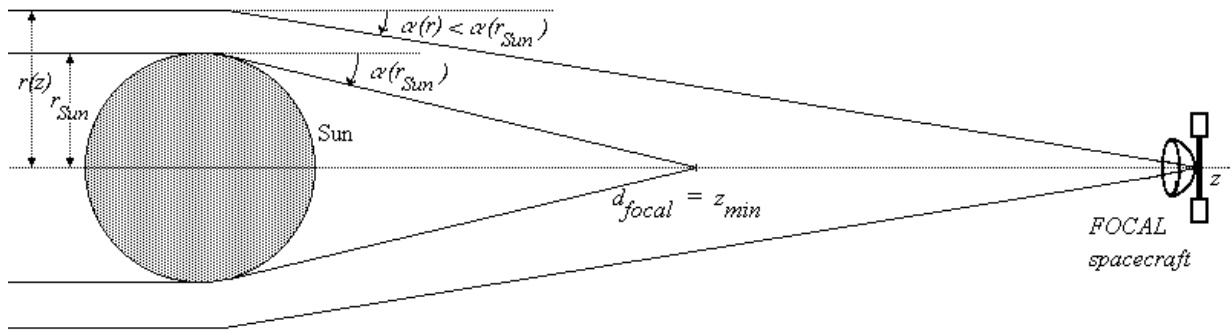

**Fig. 1.** Geometry of the Sun gravitational lens with the minimal focal length of 550 AU (= 3.17 light days = 13.75 times beyond Pluto's orbit) and the FOCAL spacecraft position beyond the minimal focal length.

The minimal focal distance (sometime rewritten by the *Schwarzschild radius* $r_{Schwarzschild} = \dfrac{2GM_{Sun}}{c^2}$ (1))

reads $d_{focal} \approx \dfrac{r_{Sun}}{\alpha(r_{Sun})} = \dfrac{c^2}{4G} \cdot \dfrac{r_{Sun}^2}{M_{Sun}} = \dfrac{r_{Sun}^2}{2r_{Schwarzschild}}$ (2). That is $d_{focal} \cong 542$ AU $\approx 550$ AU $\approx 3.171$ light days (3).

***This is the minimal distance from the Sun's center that the FOCAL spacecraft must reach to get magnified pictures of sources located on the other side of the Sun with respect to the spacecraft position.*** Also, a simple and important consequence is that ***all points on the straight line beyond this minimal focal distance are foci too***. In fact, the light rays passing by the Sun further than the minimum distance have smaller deflection angles and thus come together at an even greater distance from the Sun. Thus, ***it is not necessary to stop the FOCAL spacecraft at 550 AU. It can go on to almost any distance beyond and focalize as well or better.*** The further it goes beyond 550 AU the better it is, since the less distorted are the radio waves by the Sun Corona fluctuations. These are, in short, all FOCAL missions.

## 2. INTERSTELLAR RADIO BRIDGES FOR CHEAP INTERSTELLAR COMMUNICATIONS

In two recent published papers (refs. [2] and [3]) this author mathematically described the "radio bridges" created by the gravitational lens of the Sun and of any nearby star like Alpha Centauri A, or Barnard's star, or Sirius. The result is that it is indeed possible to communicate between the solar system and a nearby interstellar system with modest signal powers if two FOCAL mission as set up: 1) One by Humans at least at 550 AU from the Sun in the opposite direction to the selected star, and 2) one by ETs at the minimal focal distance of their own star in the direction opposite to the Sun. Open-minded readers should realize that a Civilization much more advanced than Humans in the Galaxy might already have created such a network of cheap interstellar links: a truly GALACTIC INTERNET that Humans will be unable to access as long as they won't have access to the magnifying power of their own star, the Sun, i.e. until they will be able to reach the minimal focal distance of 550 AU by virtue of their own FOCAL space missions. But let us now go back to refs. [2] and [3]. The mathematical proofs given there will not be reproduced here, since that would take too much space. The interested reader may find them in the original papers. Here we just confine ourselves to the following diagrams, showing how the Bit Error Rate (BER) of a radio bridge between, say, the Sun and Alpha Centauri A, reached the desired value of zero ("perfect communication") as long as the transmitted power $P_t$ reached the modest value of $10^{-4}$ watts for the Sun-Alpha Cen A bridge, and similarly (with higher powers) for all other bridges.

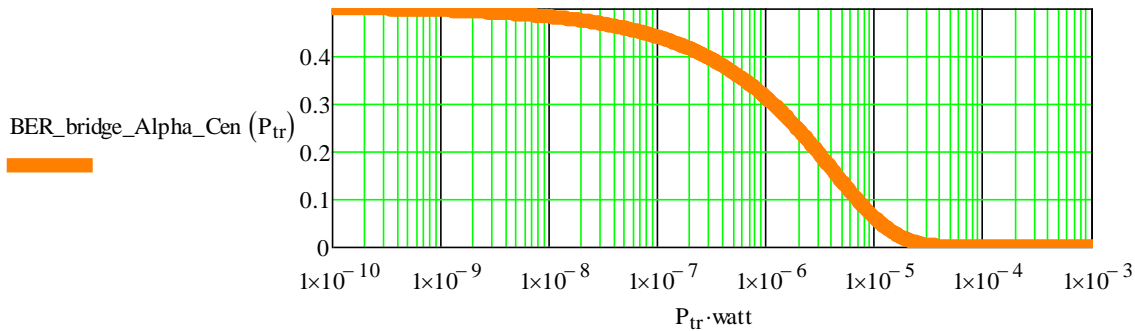

**Fig. 2.** Bit Error Rate (BER) for the double-gravitational-lens system giving the radio bridge between the Sun and Alpha Cen A. In other words, there are two gravitational lenses in the game here: the Sun one and the Alpha Cen A one, and two 12-meter FOCAL spacecrafts are supposed to have been put along the two-star axis on opposite sides at or beyond the minimal focal distances of 550 AU and 749 AU, respectively. This radio bridge has an OVERALL GAIN SO HIGH that a miserable $10^{-4}$ watt transmitting power is sufficient to let the BER get down to zero, i.e. to have perfect telecommunications! Fantastico! Notice also that the scale of the horizontal axis is logarithmic, and the trace is yellowish since the light of Alpha Cen A is yellowish too. This will help us to distinguish this curve from the similar curve for the Barnard's Star, that is a small red star 6 light years away, as we study next.

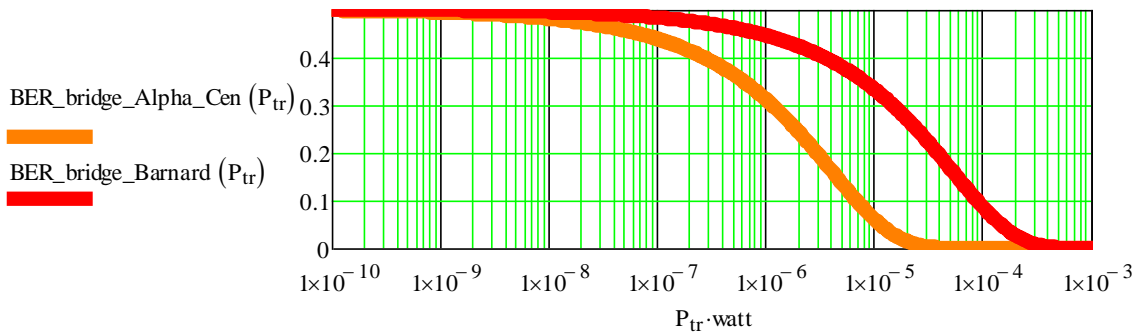

**Fig. 3.** Bit Error Rate (BER) for the double-gravitational-lens of the radio bridge between the Sun and Alpha Cen A (yellowish curve) plus the same curve for the radio bridge between the Sun and Barnard's star (reddish curve, just as Barnard's star is a reddish star): for it, $10^{-3}$ watt are needed to keep the BER down to zero, because the gain of Barnard's star is so small when compared to the Alpha Centauri A's.

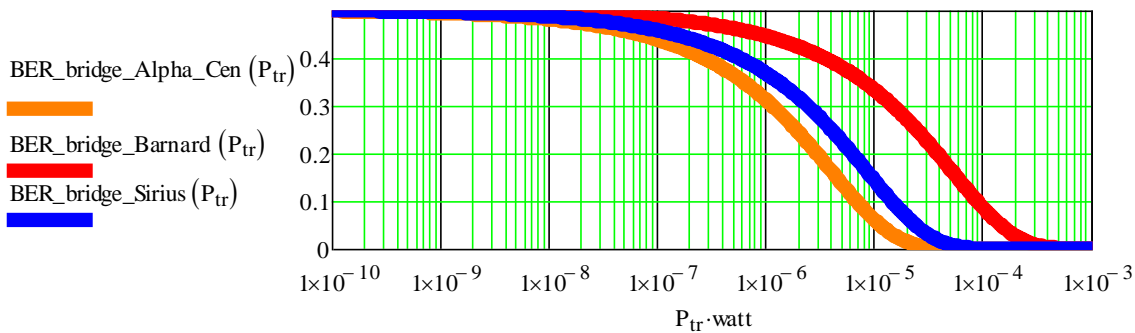

**Fig. 4.** Bit Error Rate (BER) for the double-gravitational-lens of the radio bridge between the Sun and Alpha Cen A (yellowish curve) plus the same curve for the radio bridge between the Sun and Barnard's star (reddish curve, just as Barnard's star is a reddish star) plus the same curve of the radio bridge between the Sun and Sirius A (blue curve, just as Sirius A is a big blue star). From this blue curve we see that only 10⁻⁴ watt are needed to keep the BER down to zero, because the gain of Sirius A is so big when compared the gain of the Barnard's star that it "jumps closer to Alpha Cen A's gain" even if Sirius A is so much further out than the Barnard's star! In other words, the star's gain and its size combined matter even more than its distance.

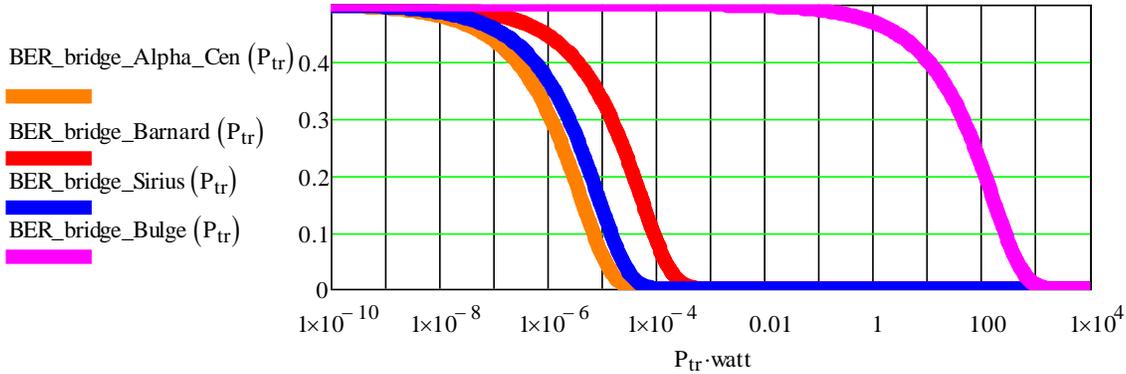

**Fig. 5.** Bit Error Rate (BER) for the double-gravitational-lens of the radio bridge between the Sun and Alpha Cen A (orangish curve) plus the same curve for the radio bridge between the Sun and Barnard's star (reddish curve, just as Barnard's star is a reddish star) plus the same curve of the radio bridge between the Sun and Sirius A (blue curve, just as Sirius A is a big blue star). In addition, to the far right we now have the pink curve showing the BER for a radio bridge between the Sun and another Sun (identical in mass and size) located inside the Galactic Bulge at a distance of 26,000 light years. The radio bridge between these two Suns works and their two gravitational lenses works perfectly (i.e. BER = 0) if the transmitted power is higher than about 1000 watts.

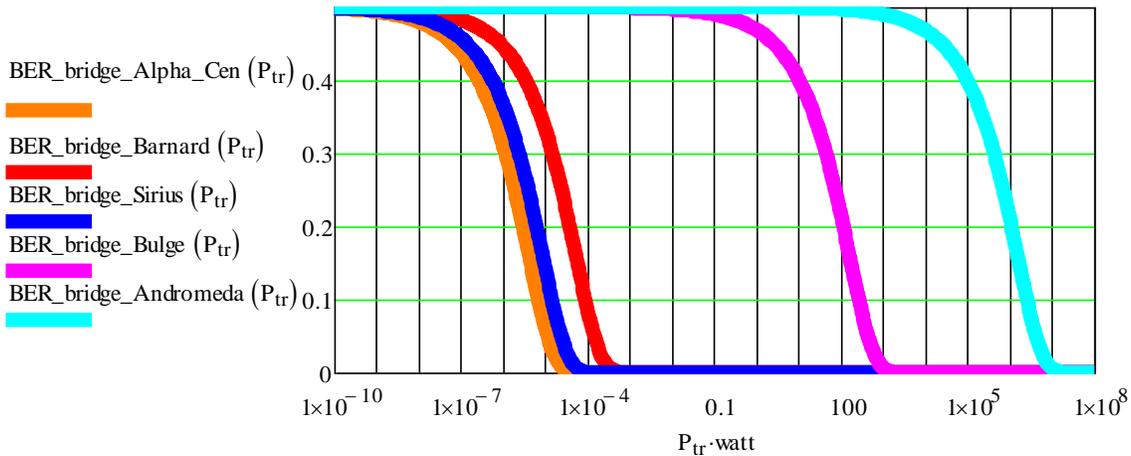

**Fig. 6.** The same four Bit Error Rate (BER) curves as shown in Figure 9 plus the new cyan curve appearing here on the far right: this is the BER curve of the radio bridge between the Sun and another Sun just the same but located somewhere in the Andromeda Galaxy M 31. Notice that this radio bridge would work fine (i.e. with BER = 0) if the transmitting power was at least $10^7$ watt = 10 Megawatt. This is not as "crazy" at it might seem if one remembers that recently (June 2009) the discovery of the first extrasolar planet in the Andromeda Galaxy was announced, and the method used for the detection was just GRAVITATIONAL LENSING !

## 3. CHANNEL CAPACITIES FOR THE INTERSTELLAR RADIO BRIDGES

The goal of this paper is to produce new results in addition to the results already proven in refs. [2] and [3]. This we now achieve by resorting to the notion of "information channel capacity" $C$ (Shannon, 1948, see the site http://en.wikipedia.org/wiki/Channel_capacity): $C = W \log_2\left(1 + \frac{P_r}{N_r}\right)$ (4), where $\frac{P_r}{N_r}$ is the signal-power-to-noise-power ratio (or simply signal-to-noise ratio, abbreviated SNR) at the receiver. Since the receiver noise is thermal, the Johnson-Nyquist formula (1928) holds $N_r = W k_b T_r$ (5) (see the site http://en.wikipedia.org/wiki/Johnson%E2%80%93Nyquist_noise ), where $k_b$ is the Boltzmann constant, $k_b = 1.3806488(13) \times 10^{-23} J/K$. Replacing (5) into (4), the information capacity of the channel made up by every radio bridge thus becomes $C(W) = W \log_2\left(1 + \frac{P_r}{W k_b T_r}\right)$ (6). On the other hand, in ref. [2], eq.(22), it was proven that $P_r = \frac{P_t A_{et} A_{er}}{r^2 \lambda^2}$ (7), where $P_t$ is the power transmitted by ETs, $A_{et}$ and $A_{er}$ are the effective areas of the transmitting and receiving antennas, respectively, $r$ is the distance between the Sun and the star around which ETs live, and $\lambda$ is the transmission wavelength. Thus, having assumed that all quantities in (7) are known, as it was done in refs. [2] and [3], one may say that the channel capacity of the radio bridge (6) is basically a function of the bandwidth $W$ only. Now, the problem of how narrow or how large the bandwidth is, indeed is a key problem for SETI. So far, the assumption of most SETI scientists has been that the bandwidth $W$ must be very narrow, i.e. of the order of 1 Hz or even less, although always above the Drake-Helou limit of 0.6 Hz to take interstellar scintillation into account. Thus, over such a narrow band, the Fourier transform is the correct mathematical tool for the extraction of the ET sinusoidal carrier out of the cosmic background noise, which is indeed white over such a narrow band. The evolution of human telecommunications on Earth over the last few decades, however, clearly shows that the trend is ***not*** toward narrower and narrower bands, but rather toward larger and larger bands! This implies that the Fourier transform may ***not*** be the best mathematical tool for SETI, and this author has campaigned for over twenty years the replacement of the good old FFT by the more modern but computationally more demanding KLT (Karhunen-Loève Transform, see, for instance, ref. [4]). The bandwidth in (6) may thus be ***any*** finite positive number, rather than just a "small" positive number. Table 1 hereafter shows the Channel Capacities given by (6) for 1 Hz and 1 kHz bandwidth for all radio bridges considered in the previous section.

| Radio Bridge | Bandwidth = 1 Hz | Bandwidth = 1 kHz | Infinite Bandwidth |
|---|---|---|---|
| Sun – Alpha Cen A | 37.6 bit/sec | 27.6 kbit/sec | 210 Gbit/sec |
| Sun – Barnard's Star | 33.9 bit/sec | 23.9 kbit/sec | 16.3 Gbit/sec |
| Sun – Sirius A | 36.5 bit/sec | 26.5 kbit/sec | 99.7 Gbit/sec |
| Sun – Sun at Bulge | 12.3 bit/sec | 2.4 kbit/sec | 5.4 kbit/sec |
| Sun – Sun at Andromeda | -0.8 bit/sec < 0 IMPOSSIBLE | -10 kbit/sec < 0 IMPOSSIBLE | 0.5 bit/sec |

**Table 1. Channel Capacities** for all five information channels made up by the radio bridges between the Sun and another star as considered in Section 2. The second and third column give the Channel Capacity for bandwidth equal to 1 Hz (typical SETI case) and 1 kHz (sometimes used in SETI also), respectively. But the "novelty" shown by this table is the last column yielding the Channel Capacities for an ***infinite*** bandwidth! This is hard to understand unless we realize that the curve $C(W)$ given by the Shannon formula (6) has a ***horizontal asymptote*** (i.e. an ***upper bound***) for $W \to \infty$, as we now prove.

Consider eq. (6) again. First of all, it is physically obvious that one must have $C(0)=0$, but the proof has to resort to L'Hospital's rule to solve the $0 \cdot \infty$ undetermined form. In other words, one has:

$$\lim_{W \to 0} C = \lim_{W \to 0}\left[W \log_2\left(1+\frac{P_r}{W k_b T_r}\right)\right] = 0 \cdot \infty = \lim_{W \to 0}\left[\frac{\log_2\left(1+\frac{P_r}{W k_b T_r}\right)}{\frac{1}{W}}\right] = \lim_{W \to 0}\left[\frac{-\frac{P_r}{W^2 k_b T_r}}{\ln(2)\left(1+\frac{P_r}{W k_b T_r}\right)} \cdot \frac{1}{-\frac{1}{W^2}}\right] = 0. \qquad (8)$$

Well, just the same limit for $W \to \infty$, rather than for $W \to 0$, reveals the horizontal asymptote of $C(W)$:

$$\lim_{W \to \infty} C = \lim_{W \to \infty}\left[W \log_2\left(1+\frac{P_r}{W k_b T_r}\right)\right] = \infty \cdot 0 = \lim_{W \to \infty}\left[\frac{\log_2\left(1+\frac{P_r}{W k_b T_r}\right)}{\frac{1}{W}}\right] = \lim_{W \to \infty}\left[\frac{-\frac{P_r}{W^2 k_b T_r}}{\ln(2)\left(1+\frac{P_r}{W k_b T_r}\right)} \cdot \frac{1}{-\frac{1}{W^2}}\right] = \frac{1}{\ln(2)} \cdot \frac{P_r}{k_b T_r},$$

(9). These are the values of the Channel Capacity for INFINITE Bandwidth shown in the last column of Table 1, and these are, of course, the *superior constraints* to the speed of transmission of information in SETI messages allowed by Physics!

**4. CONCLUSIONS: A GALACTIC INTERNET MAY EXIST ALREADY, BUT NOT FOR US**

The conclusions reached by this author in this paper are thus:
1) A Galactic Internet constructed by advanced Aliens by exploiting the gravitational lenses of stars may already exist in the Galaxy.
2) The Channel Capacity for each radio bridge between any couple of communicating stars has an upper physical limit (in bits/sec), given by $\frac{1}{\ln(2)} \cdot \frac{P_r}{k_b T_r}$, where $P_r$ is the received signal power, $T_r$ is the noise temperature of the receiving antenna (radiotelescope) and $k_b$ is Boltzmann's constant.
3) This Galactic Internet is currently inaccessible to Humans since Humans have not yet reached the minimal focal sphere of the Sun at 550 AU (and beyond, to, say, 1000 AU) by virtue of suitable FOCAL spacecrafts.
4) Even after the focal sphere of the Sun at 550 AU will have been reached by FOCAL spacecrafts, these must be aligned with the star sending the ET signals towards the Sun. Thus, the construction of some sort of "Dyson sphere for telecommunications" around the Sun at 550 AU by future, more advanced Humans might be advisable to put us in touch with the rest of the Galaxy for the first time.

**REFERENCES**
[1] C. Maccone, "Deep Space Flight and Communications – Exploiting the Sun as a Gravitational Lens", a 400-pages technical treatise published by Praxis-Springer in 2009, ISBN 978-3-540-72942-6. This book embodies all the material of the two books previously published by the author plus some fifty research papers published in peer-reviewed journals about the Sun as a Gravitational lens and the KLT in the thirty year time span between 1980 and 2009.
[2] C. Maccone, "Interstellar Radio Links Enhanced by Exploiting the Sun as a Gravitational Lens", paper #IAC-09-D4.1.8 presented at the 60th International Astronautical Congress (IAC) held in Daejeon, Republic of Korea, October 12th thru 16th, 2009. Later published in Acta Astronautica, Vol. 68 (2011), pages 76–84.
[3] C. Maccone, "Interstellar Radio Links Enabled by Gravitational Lenses of Sun and Stars", in "Communication with Extraterrestrial Intelligence", Douglas A. Vakoch (editor), State University of New York Press (2011), ISBN 978-1-4384-3793-4, pages 177-213.
[4] C. Maccone, "The KLT (Karhunen–Loève Transform) to extend SETI searches to broad-band and extremely feeble signals", Acta Astronautica, Vol. 67 (2010), pages 1427–1439.